\documentstyle[preprint,version2,aps]{revtex}

\begin{document}
\draft
\begin{title}
Alternating commensurate-incommensurate structures in the magnetic phase 
diagram of CsNiF$_3$ 
\end{title}

\author{C. Pich and F. Schwabl}
\begin{instit}
TU-M\"unchen, James-Franck-Strasse, 85747 Garching, Germany
\end{instit}
                                    
\begin{abstract}
The magnetic phase diagram of the quasi one-dimensional spinchain
system CsNiF$_3$ below the N\'eel temperature is determined. For magnetic
fields perpendicular to the spin chains incommensurate phases are predicted.  
From linear spin-wave theory we  obtain the instability line of the
paramagnetic phase as a function of the strength and the direction of the
field. The system undergoes a transition  to a commensurate or an
incommensurate phase depending on the direction of the magnetic field. In the
commensurate phase the characterizing wave vector is locked to values
describing a two-sublattice structure, whereas in the incommensurate phase the
wave vector changes continuously between the corresponding two-sublattice wave
vectors.
\end{abstract}
\narrowtext

The quasi-one-dimensional magnetic compounds ABX$_3$ (A alkaline metal, B
transition metal, X halogen) have attracted much interest
\cite{Dorner,Diep}. These materials have ferromagnetic or antiferromagnetic
intrachain interactions and weak antiferromagnetic interchain interactions 
in the plane. The systems with integer spin (e.g. CsNiCl$_3$) are studied 
in context with the Haldane conjecture \cite{Haldane}. CsFeCl$_3$ is 
discussed as an example of a system with a singlet-ground-state 
\cite{Dorner}.

CsNiF$_3$ is an example of a ferromagnet with planar
anisotropy which has been studied extensively experimentally and theoretically
\cite{STEI76,STEI91,STEI73,Shiba82a,Yamazaki80}. It remains
in the center of focus \cite{baehr,campa} because it is a model
system with reduced dimensionality where enhanced fluctuations have a
pronounced effect on the ordering structure. Due
to the planar anisotropy the spins will be oriented in the plane 
perpendicular to the chain axis. Neutron scattering revealed the dynamics 
of linear and nonlinear excitations \cite{STEI91}. 
The one dimensional spin
chain has no long range order but when applying a homogeneous field 
perpendicular to the chain axis spin waves can be measured and described 
within a linear spin wave theory \cite{STEI77}. However, the 
three-dimensional properties of CsNiF$_3$
have hardly been studied. Recently Baehr et al. \cite{baehr} measured the
magnetic excitations in the three dimensional ordered state ($T<T_N=2.7$K).
It could be shown that an isotropic antiferromagnetic exchange in the 
hexagonal plane and the dipole-dipole interaction are responsible for the 
three dimensional collinear long range order. The spins are oriented within
the plane perpendicular to the chain axis along the crystal axes of the
triangular lattice.

In the following we study the three dimensional magnetic structure of
CsNiF$_3$ for magnetic fields oriented in the plane. We examine the 
stability of the paramagnetic phase as a function of the
direction of the field. By means of linear spin wave theory we obtain the
instability line at which the paramagnetic spin orientation gets unstable 
and changes to a canted structure. It turns out that the ground state 
depends sensitively on the field direction. For certain angular domains
($\Delta\varphi_{ic}\approx 15.6^o$) there exist incommensurate phases
separated by commensurate phases (of width $\Delta\varphi_{c}\approx
44.4^o$). Thus as a function of the direction of the field we obtain
alternating commensurate and incommensurate spin structures. This is the 
first time that incommensurate phases have been predicted in CsNiF$_3$. 

The starting point of our investigation is the Spin Hamiltonian 
\begin{equation}
  H = -2J\sum_{i}{\bf S}_i{\bf S}_{i+1}+A\sum_{l} {(S_l^z)}^2 -
  \sum_{\alpha,\beta}\sum_{l,l'}\left(J_{ll'}'\delta^{\alpha\beta}+ 
    A^{\alpha\beta}_{ll'}\right)S_l^\alpha S_{l'}^\beta -g_L\mu_B {\bf
    H}_0\sum_l {\bf S}_l \ . 
  \label{hamilton}
\end{equation}
Here $J$ denotes the ferromagnetic nearest-neighbor intrachain 
interaction, $A$ the single-ion anisotropy, $J_{ll'}'$ the interchain 
and $i$ indicates positions on one and the same spin chain, whereas 
$l$ indicates 
all spin positions. ${\bf H}_0$ is an external field perpendicular to 
the chain
axis. The magnetic lattice structure of CsNiF$_3$ is a simple hexagonal
structure with lattice constants $c=2.6$ \AA{} of the spin chains and 
$a=6.2$ \AA{} of the triangular lattice in the plane.
From neutron scattering \cite{baehr} the coefficients in Eq. 
(\ref{hamilton}) are deduced to $J=11.8$ K, $A=3.3$ K and $J'=-0.025$ K. 
The value for $J'$ is of the same order as the dipole energy which is \
necessary for stabilizing a collinear antiferromagnetic spin structure. 
Because of the large planar anisotropy the spins are forced to lie in the
hexagonal plane. The classical 
ground state for vanishing field is given by three possible domains A--C 
(s. Fig. \ref{Spinstruct0}) in which a collinear antiferomagnetic spin 
structure is realized, i.e. the rotation symmetry is broken due to the
competition of the exchange and the dipole-dipole interaction
\cite{baehr,pich96}. Thus the system has an easy axis anisotropy. In the 
real system all three domains are simultanously present.

Fourier transformation of the Hamiltonian (Eq. (\ref{hamilton})) yields  
\begin{equation}
  H = -\sum_{\alpha,\beta}\sum_{\bf q} \left( J_{\bf q}\delta^{\alpha\beta}
-  A\delta^{\alpha z}\delta^{\beta z}+J_{\bf q}'\delta^{\alpha\beta}+
    A^{\alpha\beta}_{\bf q}\right)S_{\bf q}^\alpha S_{-{\bf q}}^\beta 
  -g_L\mu_B\sqrt N {\bf H}_0{\bf S}_0\, ,
\label{HH}
\end{equation}
with the nearest-neighbor exchange energies (intrachain and interchain)
\begin{eqnarray}
  J_{\bf q} & = & 2J\cos cq_z \\ J_{\bf q}'& = & 2J'\left(\cos aq_x
    +2\cos{aq_x\over 2}\cos{\sqrt 3aq_y\over 2}\right) \, .
\end{eqnarray}
$A^{\alpha\beta}_{\bf q}$ denotes the Fourier transform of the long-range
dipole interaction, which is calculated by means of the Ewald summation
technique \cite{Bonsal77,Keffer55}. 

The qualitative effect of a homogeneous magnetic field is as follows.
For weak magnetic fields transverse to the spin orientation the spins will
reorient in order to gain energy from the Zeeman term. Thus when orienting
the field along one domain direction the two others will change their 
ground state immediately. For a field longitudinal to the spins a spin-wave
calculation reveals \cite{pich96,Pich93} that the N\'eel state is at least
metastable up to a finite critical field which depends solely on
the dipole energy. Thus the situation for a virgin probe neglecting effects
from domain wall energies and crystal defects is as follows: Without 
magnetic field the system might be built up of the three domains in equal
fractions. Raising the magnetic field (parallel to the spins in domain A) 
does not change the spin orientation in domain A but leads to a slight 
reorientation in domains B and C. Above the the critical value the spins 
in domain A flip (first order phase transition) to an orientation 
identical to either domain B 
or domain C. For strong magnetic fields one finally enters the paramagnetic
phase. When the magnetic field is decreased thereafter, the spins order 
again in the two domains but domain A is not formed any more because of the
metastability of this domain. The system ends up in a state, where only 
domains B and C are present.  

In the following we study the excitations by means of linear spin wave 
theory. Via a Holstein-Primakoff-transformation we transform the Spin 
operators
$S_l^\alpha$ in the Hamiltonian (Eq. \ref{HH}) to Bose operators $a_l$ and
$a_l^{\dag}$ \cite{Ziman69,Keffer66} and diagonalize
the quadratic form. 

In the paramagnetic phase all spins are aligned along the magnetic field 
${\bf H}_0$. Because the ground state of CsNiF$_3$ for vanishing fields is 
not invariant with respect to a rotation around the spin-chain axis 
(recall that
there are three domains A--C), the direction of the field plays a crucial
role. The field direction is parametrized by $\varphi$, which is the angle
between the field and the $x$-axis (inset in Fig. (4)). To quadratic order 
in the Bose operators we obtain the Hamiltonian 
\begin{equation}
  H  = E_{PM}^{cl}+ \sum_{\bf q}A_{\bf q}~a_{\bf
    q}^{\dagger}a_{\bf q} + {1\over 2}\left( 
    B_{\bf q}a_{\bf q}~a_{-\bf q}+B_{\bf q}^\ast a_{\bf q}^{\dagger}a_{-\bf
    q}^{\dagger}\right)
\end{equation}
with the coefficients
\begin{eqnarray}
A_{\bf q}  & = & S(2(J_0-J_{\bf q})+A+2J_0'-2J_{\bf q}')+g_L\mu_B H_0 
\nonumber
\\ 
&&  +S\left(\cos^2\varphi (2A^{xx}_0 -A^{yy}_{\bf q})+\sin^2\varphi
(2A^{yy}_0-A^{xx}_{\bf q}) -A^{zz}_{\bf q} -
\sin{2\varphi} A^{xy}_{\bf q}\right) \\
B_{\bf q}  & = & S(-A+A^{zz}_{\bf q}-\sin^2\varphi A^{xx}_{\bf
q}-\cos^2\varphi A^{yy}_{\bf q}+\sin{2\varphi} A^{xy}_{\bf q})\nonumber
\end{eqnarray}
and the classical ground state energy of the paramagnetic state
\begin{equation}
E_{PM}^{cl} = -NS^2\left(J_0+J_0' +(\cos^2\varphi A^{xx}_0+\sin^2\varphi
A^{yy}_0)\right) 
-g_L\mu_BNSH_0\, ,
\end{equation}
which is independent of the direction of the field for spherical shaped
systems ($A_0^{xx}=A_0^{yy}$). Here we considered only  wave vectors within
the plane ($q_z=0,A_{\bf q}^{yz}=0 $) due to the strong planar anisotropy. 
The dispersion relation for the paramagnetic phase is calculated via a 
Bogoliubov transformation to 
\begin{equation}
E_{\bf q} = \sqrt{A_{\bf q}^2-|B_{\bf q}|^2}\, .
\label{Eqq}
\end{equation}
This equation is valid for high magnetic fields. When lowering the magnetic
field the paramagnetic phase gets unstable and changes to a canted spin
structure. This instability is signalled by a soft mode at the wave vector
${\bf q}(\varphi)$ characterizing the phase below the paramagnetic phase. 
The
value of the critical field $H_0^c$ and the wave vector ${\bf q}(\varphi)$
depend on the angle of the field $\varphi$ and can be evaluated from
Eq. (\ref{Eqq}) by setting the excitation energy to zero. This leads to the
equation 
\begin{equation}
  g_L\mu_BH_0^c(\varphi) = 2S \left(J_{\bf q}'-J_0'+\sin^2\varphi 
(A^{xx}_{\bf
  q}-A^{yy}_0)+ \cos^2\varphi (A^{yy}_{\bf q}-A^{xx}_0) -\sin{2\varphi} 
  A^{xy}_{\bf q} \right )\, .
\label{H_0}
\end{equation}
Note that this expression is independent of the ferromagnetic exchange
and the anisotropy energies as long as they are much larger than the
antiferromagnetic exchange and the dipole energies. Regarded as a function
of ${\bf q}$ the maximum of this expression gives the angular dependent
critical value $H_0^c(\varphi)$ and the wave vector ${\bf
q}(\varphi)$. Assuming that the phase joining the paramagnetic is a
conventional spin-flop phase ($\alpha=\beta$ in inset of Fig. (4)) or a 
general two-sublattice structure, the so-called intermediate phase 
($\alpha\ne\beta$ in inset of Fig. (4)), we expect the maximum value for 
the critical field at wave vectors describing the antiferromagnetic 
domains, i.e. ${\bf q}_1={2\pi\over
\sqrt 3a}(0,1,0)$, ${\bf q}_2={\pi\over a} (1,1/\sqrt 3,0)$ or ${\bf
q}_3={\pi\over a}(1,-1/\sqrt 3,0)$ (s. Fig. 2). However, the detailed 
analysis
shows that the paramagnetic phase gets unstable at an incommensurate wave
vector ${\bf q}(\varphi)$ for certain field directions. Before considering 
the general case we study the special angles $\varphi =0^o$ and 
$\varphi=90^o$.

(i) $\varphi =90^o$: For magnetic fields parallel to the $y$-axis
the critical value is given by 
\begin{equation}
  g_L\mu_BH_0^{cy} = 2S (J_{\bf q}'-J_0'-A^{yy}_0+ A^{xx}_{\bf q})\, .
\end{equation}
This expression is found to have its maximum value of at ${\bf q}_1$,
the antiferromagnetic wave vector of the collinear phase of domain A. The
actual value for the critical field is evaluated with the parameters given
above to (spherical shape)
\begin{equation}
H_0^{cy} = 340 {\rm mT} \qquad {\rm at} \qquad {\bf q}(90^o) = {\bf q}_1
\, .
\end{equation}
At $H_0^{cy}$ the system undergoes a transition into a commensurate phase,
precisely to the two sublattice phase described by ${\bf q}_1$. Owing
to the hexagonal symmetry, the critical field for $\varphi =30^o$ is the 
same
as for $\varphi =90^o$ but at the antiferromagnetic wave vector ${\bf q}_2$
characterizing domain C; for $\varphi =-30^o$ the structure is given by 
${\bf q}_3$ corresponding to domain B (s. Fig. (5)).

(ii) $\varphi =0^o$: For this field direction Eq. (\ref{H_0}) reduces to: 
\begin{equation}
  g_L\mu_BH_0^{cx} = 2S (J_{\bf q}'-J_0'-A^{xx}_0+ A^{yy}_{\bf q})\ .
\end{equation}
The maximum value is not achieved for any of the two-sublattice wave 
vectors
but for a wave vector with only an ${\bf q}_x$-component depending on the relative
strength of the antiferromagnetic exchange and the dipole energy. This 
follows
from the fact that the dipole component $A^{yy}_{\bf q}$ has a linear wave
vector dependence at ${\bf q}_0$ rather than a quadratic as found for 
$J_{\bf
q}$ \cite{Shiba82}. Evaluation of the critical value for CsNiF$_3$ leads to
(spherical shape) 
\begin{equation}
H_0^{cx} = 290 {\rm mT} \qquad {\rm at} \qquad {\bf q}(0) = {\pi\over
a}(1.023,0,0)\, .
\end{equation}
Thus the system undergoes a transition to an incommensurate phase. The
incommensurate wave vector happens to be near the wave vector ${\bf 
q}_4={\pi\over a}(1,0,0)$, characterizing a four-sublattice structure, 
which describes an antiferromagnetic modulation along the  $x$-axis.

(iii) arbitrary $\varphi=$: Finally we turn to arbitrary angles, for which 
the 
situation turns out to be nontrivial. The complete dependence on the field
direction ${\bf q} (\varphi)$ for CsNiF$_3$ is plotted in
Fig. (\ref{inkomm-wave}). The wave vector components $q_x$ and $q_y$ are
plotted as a function of the angle $\varphi$ for the region in question. A
critical angle $\varphi_c \approx 7.8^o$ is obtained above which 
the wave vector is locked to ${\bf 
q}_2$. Varying the angle $\varphi$ between $0^o$ and $7.8^o$ the wave 
vector
where the instability sets in changes continuously from ${\bf q}(0)$ to
${\bf q}(\varphi_c)={\bf q}_2$. In Fig. (\ref{inkomm-wave-H_0}) the 
critical value for the magnitude of the magnetic field is plotted as a 
function
of the angle for the same angular domain. Note that the critical field is
continuous even at the critical angle $\varphi_c$. It has only a small kink
at this point. The dashed curve results from Eq. (\ref{H_0}) under the 
assumption
that the instability point occurs at ${\bf q}_2$ for the whole angular
segment. Herefrom we can see that for $\varphi < \varphi_c$ the 
incommensurate
structure is favored. Due to the inversion symmetry of the lattice together
with ${\bf q}(\varphi)$ there is a second modulation wave vector
$-{\bf q}(\varphi)$. The result of this investigation for all angles
$\varphi$ of the magnetic field is summarized in Fig. (\ref{two-inkom}): 
The instability line of the paramagnetic phase is shown. There are angular
regions 
where the paramagnetic structure undergoes a transition to a two-sublattice
structure (annotated by the corresponding wave vector) seperated by regions
drawn in thick where an incommensurate structure is formed.

Examination of the commensurate phase leads to the following results: The
commensurate phases are described near the transition to the paramagnetic 
phase by a two-sublattice structure. Calculation of the classical ground 
state energy \cite{pich96} shows that a conventional spin-flop phase which 
is parametrized
by a single angle for both sublattice spins is stable only for the field
directions of $\varphi =30^o+n60^o$ and integer values of $n$. For all 
other field directions the commensurate phase is an intermediate phase with
two independent angles.

In summary we have studied the ground state for CsNiF$_3$ with a 
homogeneous magnetic field oriented in the hexagonal plane. By stability 
investigations of the paramagnetic phase we obtained a non-circular 
instability line for fields
in the hexagonal plane. The magnitude of the critical field and the type of
phase joining the paramagnetic phase depends crucially on its direction.
For certain angular domains the system changes to incommensurate
structures, which are seperated by commensurate (two-sublattice) ones
(Fig. (\ref{two-inkom})). This is the first time a staircase like behavior 
is found in a magnetic system as a function of the direction of the 
field. The number of steps is 6, which corresponds to twice the number of
two-sublattice wave vectors. This staircase is quite different from a 
devil's staircase with an infinite number of lock-in steps 
\cite{bak,Selke}.  Our results for CsNiF$_3$ are in contrast to Yamazaki et
al.\cite{Yamazaki80}, 
who predicted conventional spin-flop phases for all field directions.
This may result from their semiclassical model which does not 
consider the full nature of the dipole-dipole interaction. 
According to our theory the magnetic phase diagram shows a much richer
structure including intermediate spin configurations and incommensurate 
phases. The observation of our predicted new phases and the wave vector
dependence is left to future experiments, e.g. neutron scattering.

\acknowledgments This work has been supported by the German Federal
Ministry for Education and Research (BMBF) under the contract number
03-SC4TUM.

\figure{\label{Spinstruct0}The ground state for CsNiF$_3$ in the
  hexagonal plane is one of the three shown configurations (A--C),
  called domains. In domain A the two primitive vectors are
  represented. The antiferromagnetic modulation can be described by ${\bf
  q}_1$, ${\bf q}_2$ and ${\bf q}_3$ for domain A, B and C respectively. } 

\figure{\label{BrillFig}The Brillouin zones of the hexagonal plane. The
  hexagon is the crystallographic and the rectangle (dashed) the
  magnetic one.}

\figure{\label{inkomm-wave} Wave vector $(q_x,q_y)$ at which the
  paramagnetic phase gets unstable versus the
  direction of the external magnetic field (solid line: $q_y$, dashed line:
  $q_x$). Parameters for CsNiF$_3$ are used. For $\varphi > 7.8^o$ the
  instability appears for {\bf q}$_2$.} 

\figure{\label{inkomm-wave-H_0} Critical field below which the paramagnetic
  phase gets unstable for field direction between $0^o$ and $30^o$. For 
  smaller
  (larger) angles the paramagnetic phase changes to an incommensurate
  (commensurate) phase. The dashed curve indicates the critical field when
  assuming that the soft mode occurs at ${\bf q}_2$. The inset shows the
  coordinate system for a general two-sublattice spin orientation.}

\figure{\label{two-inkom} Angular dependence of the instability of the
  paramagnetic phase for CsNiF$_3$. The thick segments on the instability 
  curve
  correspond to directions of the magnetic field ${\bf 
  H}_0 = {\rm H}_0(\cos\varphi,\sin\varphi,0)$ for which the paramagnetic
  phase changes to an incommensurate phase. These alternate with segments 
  where
  a transition to a commensurate structure appears. The corresponding wave
  vector is given.}  


\begin{references}


  \bibitem{Dorner} B. Schmid, B. Dorner, D. Petitgrand, P.L. Regnault and
  M. Steiner, Z. Phys. {\bf B 95}, 13 (1994); P. Lindg\aa rd, B. Schmid,
  Phys. Rev. {\bf B48}, 13636 (1993).

  \bibitem{Diep} B.D. Gaulin, in {\it Magnetic systems with competing 
      interactions}, edited by H.T. Diep, World Scientific (1994).

  \bibitem{Haldane} F.D.M. Haldane, Phys. Rev. Lett. {\bf 50}, 1153 (1983);
  T. Inam, K. Kakurai, H. Tanaka, M. Enderle and M. Steiner,
  J. Phys. Soc. Jpn. {\bf 63}, 1530 (1995).

  \bibitem{STEI76} M. Steiner, J. Villain and C.G.  Windsor, Adv. Phys.
  {\bf 25}, 87 (1976).

  \bibitem{STEI91} M. Steiner M.J.  Mikeska, Adv. Phys {\bf 40}, 191 (1991).

  \bibitem{STEI73} M. Steiner, B. Dorner, Solid State Commun. {\bf 12}, 537 
  (1973); M. Steiner, H. Dachs, Solid State Commun.  {\bf 14}, 841 (1974).

  \bibitem{Shiba82a} H. Shiba, N. Suzuki, J. Phys. Soc. Jpn. {\bf 51}, 3488
  (1982).

  \bibitem{Yamazaki80} H. Yamazaki, E. Soares, H. Panepucci and Y. 
  Morishige, J. Phys. Soc. Jpn. {\bf 47}, 1464 (1979) and
  J. Phys. Soc. Jpn. {\bf 48}, 1453 (1980) 

  \bibitem{STEI77} M. Steiner, J. K. Kjems, J. Phys. C: Solid State Phys., 
    {\bf 10}, 2665 (1977).

  \bibitem{baehr} M. Baehr, M. Winkelmann, P. Vorderwisch, M. Steiner, C. 
  Pich, F. Schwabl, accepted for publication in Phys. Rev. {\bf B}.

  \bibitem{campa} L.S. Campana, A. Caramico D'Auria, F. Esposito, 
  U. Esposito and G. Kamieniarz, Phys. Rev. {\bf B 53}, 2594 (1996).


  \bibitem{pich96} C. Pich and F. Schwabl, in preparation.

  \bibitem{Bonsal77} L. Bonsal and A.A. Maradudin, Phys. Rev. B {\bf
    15}, 1959 (1977).

  \bibitem{Keffer55} M.H. Cohen, F. Keffer, Phys. Rev. {\bf 99}, 1135 
    (1955).

  \bibitem{Pich93} C. Pich and F. Schwabl, Phys. Rev. B 47, 7957
  (1993).

  \bibitem{Ziman69} J.M. Ziman, {\it Principles of the theory of
    Solids}, (Cambridge Press, 1969), p. 317ff

  \bibitem{Keffer66} F. Keffer, in {\it Encyclopedea of Physics}, 
    Vol. XVIII/2,
    edited by S. Fl\"ugge (Springer-Verlag, Heidelberg, 1966), p. 37ff

  \bibitem{Shiba82} H. Shiba, Solid State Commun.  {\bf 41}, 511
  (1982).

  \bibitem{bak} P. Bak, Rep. Prog. Phys. {\bf 45}, 587 (1982).

  \bibitem{Selke} W. Selke, in {\it Phase Transitions and Critical
  Phenomenona}, Vol. 15, eds. C. Domb and J.L. Lebowitz 
  (Academic Press, New York, 1993).
 
\end{references}
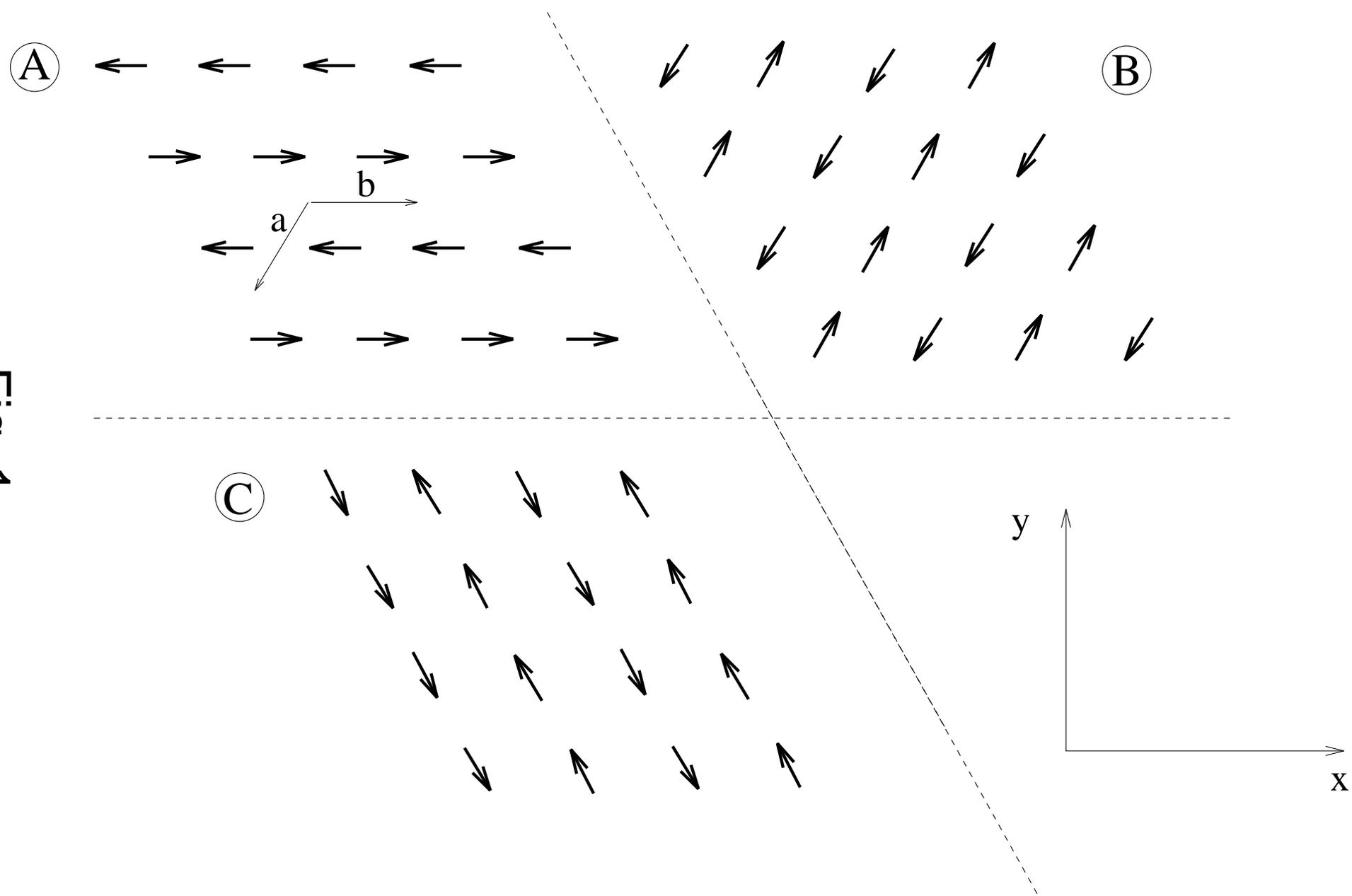
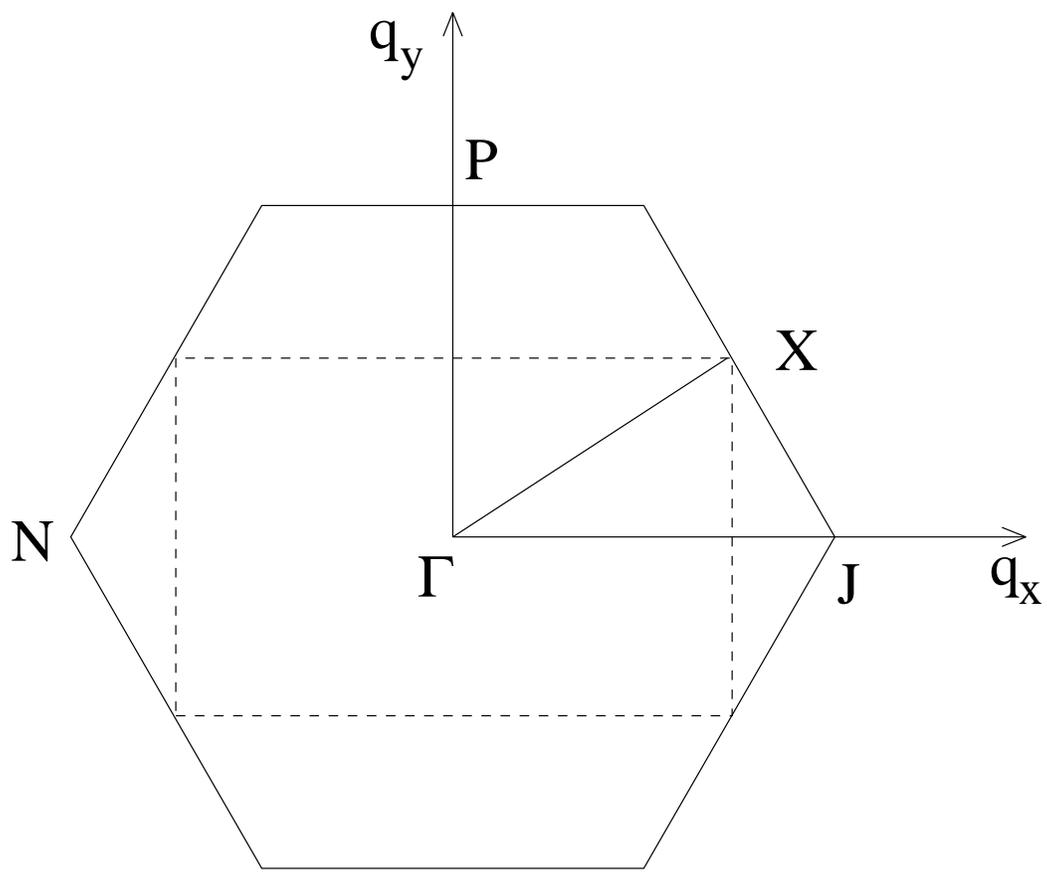
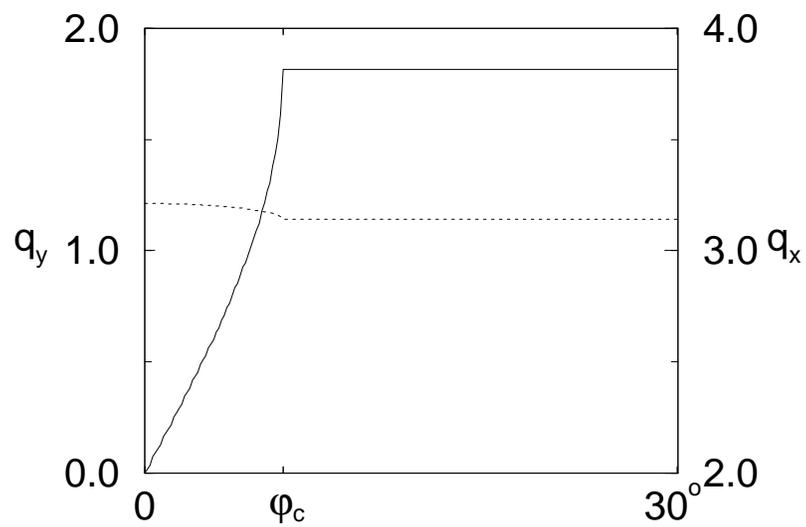
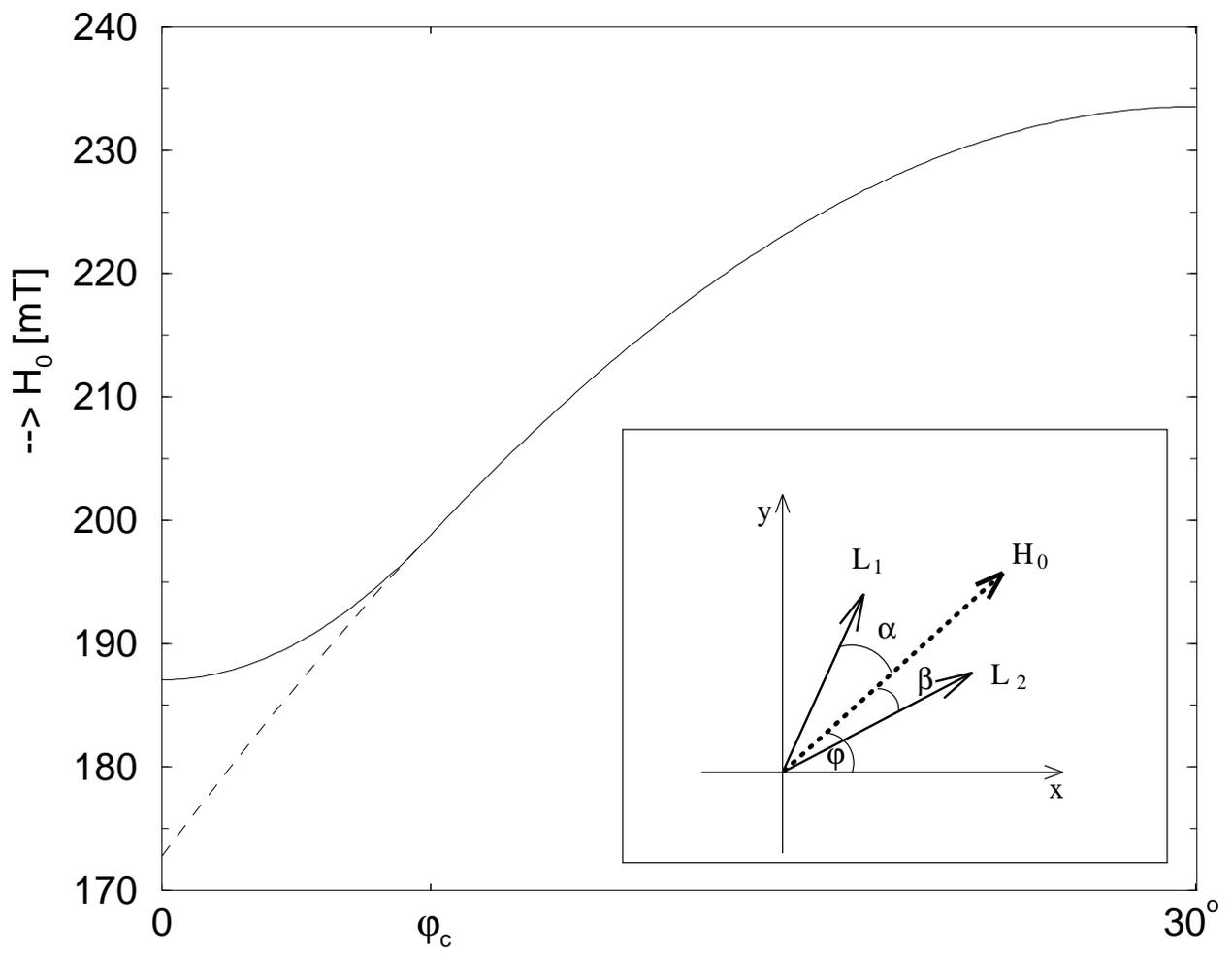
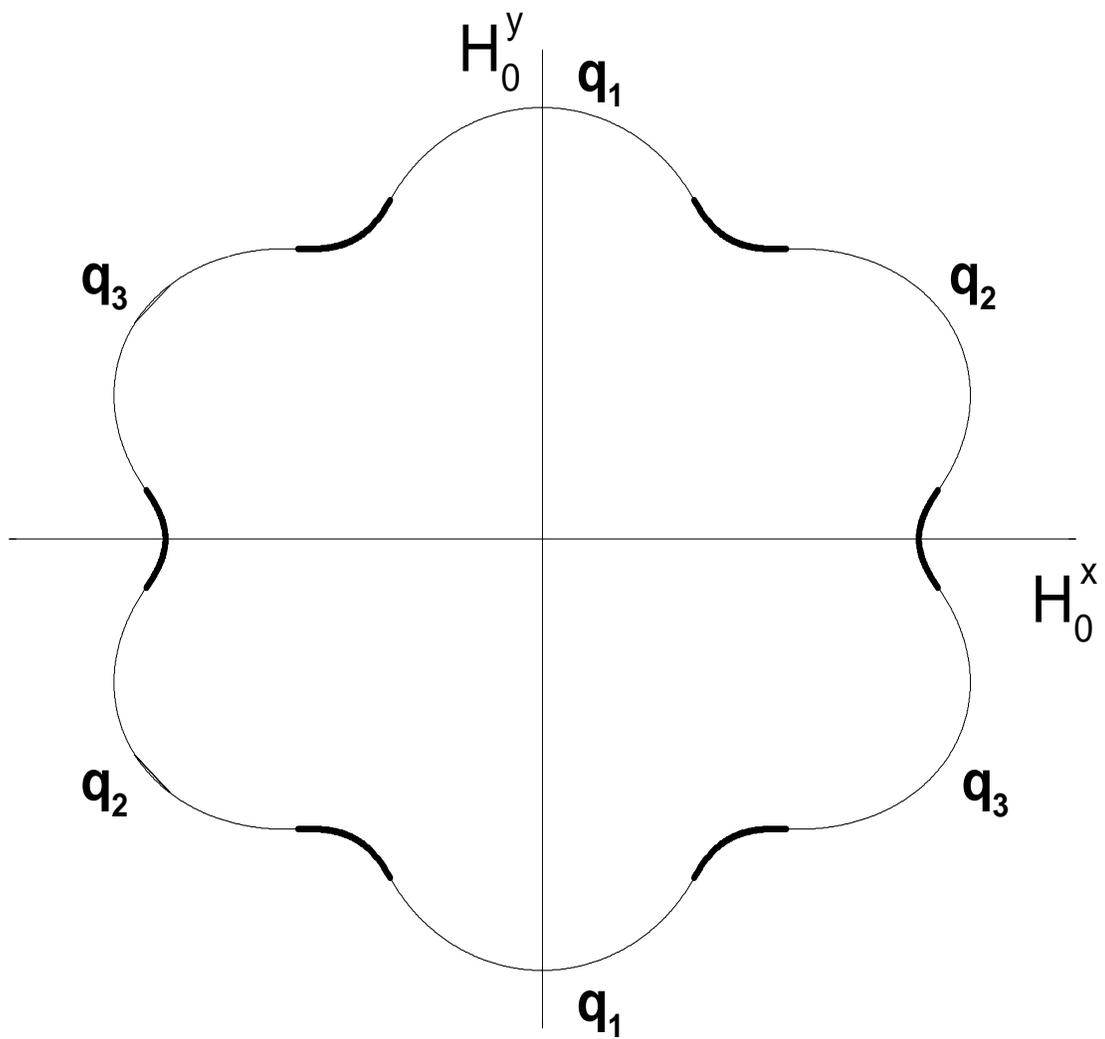
\end{document}